\documentclass[12pt,preprint]{aastex}
\usepackage[]{natbib}
\usepackage{graphics}
\usepackage{emulateapj5}
\usepackage{apjfonts}

\newcommand{\etal}{et al.}  
\newcommand{\per}{\ensuremath{^{-1}}}
\newcommand{\persq}{\ensuremath{^{-2}}}

\newcommand{\ebv}{\ensuremath{E(B-V)}}
\newcommand{\hete}{\emph{HETE-2}}
\newcommand{\tjet}{\ensuremath{t_\mathrm{jet}}}

\slugcomment{To appear in The Astrophysical Journal Letters} 
\shorttitle{SPECTROPOLARIMETRY OF GRB 020813}

\shortauthors{BARTH ET AL.}

\begin{document} 

\title{Optical Spectropolarimetry of the GRB 020813 Afterglow}

\author{
  Aaron J. Barth\altaffilmark{1,2},
  Re'em Sari\altaffilmark{3}, 
  Marshall H. Cohen\altaffilmark{1}, 
  Robert W. Goodrich\altaffilmark{4}, 
  Paul A. Price\altaffilmark{1,5}, 
  Derek W. Fox\altaffilmark{1}, 
  J. S. Bloom\altaffilmark{6}, 
  Alicia M. Soderberg\altaffilmark{1},
  and Shrinivas R. Kulkarni\altaffilmark{1}
}

\altaffiltext{1}{Astronomy Department, 105-24 Caltech, Pasadena, CA
91125; barth@astro.caltech.edu} 

\altaffiltext{2}{Hubble Fellow}

\altaffiltext{3}{Theoretical Astrophysics, 130-33 Caltech, Pasadena,
  CA 91125}

\altaffiltext{4}{W. M. Keck Observatory, 65-1120 Mamalahoa Highway,
  Kamuela, HI 96743}

\altaffiltext{5}{Research School of Astronomy and Astrophysics, Mount
  Stromlo Observatory, via Cotter Road, Weston, ACT 2611, Australia}

\altaffiltext{6}{Harvard-Smithsonian Center for Astrophyics, 60 Garden
  St., Cambridge, MA 02138}

\begin{abstract}

The optical afterglow of gamma-ray burst 020813 was observed for 3
hours with the LRIS spectropolarimeter at the Keck-I telescope,
beginning 4.7 hours after the burst was detected by \hete.  The
spectrum reveals numerous metal absorption lines that we identify with
two systems at $z=1.223$ and $z=1.255$.  We also detect an
[\ion{O}{2}] $\lambda3727$ emission line at $z=1.255$ and we identify
this galaxy as the likely host of the GRB.  After a correction for
Galactic interstellar polarization, the optical afterglow has a linear
polarization of 1.8--2.4\% during 4.7--7.9 hours after the burst.  A
measurement of $p = 0.80\% \pm 0.16\%$ on the following night by
Covino \etal\ demonstrates significant polarization variability over
the next 14 hours. The lack of strong variability in the position
angle of linear polarization indicates that the magnetic field in the
jet is likely to be globally ordered rather than composed of a number
of randomly oriented cells.  Within the framework of afterglow models
with collimated flows, the relatively low observed polarization
suggests that the magnetic field components perpendicular and parallel
to the shock front are only different by about 20\%.

\end{abstract}

\keywords{gamma rays: bursts --- polarization}

\section{Introduction}

It is now generally assumed that the optical afterglows of gamma-ray
bursts (GRBs) arise from synchrotron emission in the post-shocked gas
of a relativistic blast wave \citep[see, e.g.,][]{pir00, mes02}.
There is credible evidence that GRBs are strongly collimated or beamed
explosions \citep[e.g.,][]{rho99,sph99,fra01}.  Synchrotron radiation
is intrinsically polarized, and the synchrotron emission from a jet
will have a net polarization if its axis does not coincide with the
line of sight.  A natural consequence of models for beamed afterglows
is time-variable polarization, which will depend on the viewing
geometry as well as the degree of orientation of the magnetic field
\citep[e.g.,][]{ml99, gru99}, up to a maximum polarization of
$\sim20\%$ \citep{gl99,sari99}.  Optical imaging polarimetry has been
performed for a handful of GRB afterglows to search for this
signature.  The results have been either upper limits \citep{hjo99,
cov02a} or detections at $p \approx 1-3\%$ \citep{cov99, wij99, rol00,
cov02b}, except for an observation of GRB 020405 by \citet{ber02} that
detected $9.9\% \pm 1.3\%$ polarization 1.3 days after the burst.

GRB 020813, a typical long-duration burst with a series of pulses, was
detected by \hete\ at 2:44 UT on 2002 August 13 \citep{vil02}, and an
associated optical transient (OT) was promptly identified
\citep{fbp02}.  The rapid localization allowed us to begin
spectropolarimetric observations less than 5 hours after the burst was
initially detected.  These are the first spectropolarimetric
observations obtained for a GRB afterglow.  Preliminary descriptions
of this dataset were given by \citet{pri02} and \citet{bar02}.

\section{Observations and Reductions}

The observations were obtained with the dual-beam LRIS
spectropolarimeter \citep[LRISp;][]{oke95, gcp95} on the Keck-I
telescope, and a slit width of 1\arcsec.  The addition of a new
ultraviolet polarizing calibration filter to LRISp allowed us for the
first time to take advantage of the full available wavelength range of
LRISp using both the blue and red cameras.  We used a 400 lines
mm\per\ grism on the blue side of the spectrograph and a 400 lines
mm\per\ grating on the red side, giving a wavelength scale of 1.09
\AA\ pixel\per\ over 3200--5800 \AA\ in the blue and 1.86 \AA\
pixel\per\ over 5600--9400 \AA\ in the red.  The red and blue beams
were separated with a dichroic beamsplitter having 50\% transmission
at 5700 \AA.

We began exposures of the OT associated with GRB 020813 on 2002 August
13 at 7:23 UT, when its brightness was $V \approx 19$ mag. The total
exposure time was 3 hours, broken into individual 15-minute exposures.
Each pair of consecutive exposures gives one of the Stokes parameters
$q$ (with the half-wave plate oriented at 0\arcdeg\ and 45\arcdeg) or
$u$ (22\fdg5 and 67\fdg5), and the 4 exposures taken together are used
to determine the degree and position angle of linear polarization ($p$
and $\theta$).  We refer to these three sets of four exposures each as
the ``hour 1'', ``hour 2'', and ``hour 3'' data.  The airmass was
1.3--1.5 during the observations, and before starting each hour's
exposures the spectrograph was rotated to place the slit at the
parallactic angle for the midpoint of the hour.  Seeing was
$\sim0\farcs7$, and there were intermittent thin clouds during the
observations.

As a probe of the Galactic interstellar polarization (ISP), we also
observed the A2V star HD 187330, which is located 64\farcm7 from the
line of sight to the GRB.  Its apparent magnitude ($V = 9.22$ mag)
implies a distance of $\sim350$ pc; this is sufficiently distant to
probe the majority of the Galactic ISP for a Galactic latitude of
$-21\arcdeg$ \citep{tran95}.  Distant A-type stars are useful as
probes of ISP because of their brightness and low intrinsic
polarization, although ideally a smaller angular separation from the
OT would be desirable.

The data were reduced using standard techniques as described by
\citet{mrg88}.  The zeropoint of $\theta$ was calibrated with the
polarized standard star BD+59\arcdeg389 \citep{sel92}.  For the
unpolarized standard star HD 14069 we found $p \leq 0.1\%$ across the
entire optical spectrum.

\section{Continuum and Absorption Lines}

Figure 1 displays the total flux spectrum of the OT from all exposures
combined.  Numerous metal absorption lines are superposed on a
featureless continuum, as is typical for GRB afterglow spectra
\citep[e.g.,][]{met97, kul99, vre01, jha01}.  Two distinct absorption
systems are present in the spectrum (Table 2).  The individual
absorption lines in the higher-redshift system have a mean $z =
1.255$.  The second system has weaker absorption lines and a mean
redshift of 1.223.  Both systems show lines of \ion{Si}{2}
$\lambda1527$, \ion{C}{4} $\lambda\lambda1548, 1551$, \ion{Al}{3}
$\lambda1671$, \ion{Mg}{2} $\lambda\lambda2796, 2804$, and \ion{Mg}{1}
$\lambda2853$, as well as several \ion{Fe}{2} lines including
$\lambda\lambda2587, 2600$. Additional features from the $z=1.255$
system include lines of \ion{Al}{3}, \ion{Mn}{2}, \ion{Mg}{1},
\ion{Zn}{2}, and \ion{Cr}{2}.  One emission line is also present in
the spectrum, centered at 8404.0 \AA\ and consistent with [\ion{O}{2}]
$\lambda3727$ at $z = 1.255$.  We conclude that the $z=1.255$ system
is the likely host of the GRB, since it is the highest-redshift
absorption system detected and its [\ion{O}{2}] emission line is an
indicator of star-forming activity.

Ionization of the surrounding interstellar medium (ISM) by the GRB
could cause temporal variations in the equivalent widths of
interstellar absorption lines within the host galaxy \citep{pl98}, and
we have measured the equivalent widths of the absorption lines in the
hour 1, 2, and 3 spectra separately to search for any variations.  The
differences in equivalent widths are consistent with the level of
measurement uncertainty, and we conclude that no variability is
detected.  Searches for ultraviolet absorption-line variability in
other GRBs have so far yielded null results as well \citep{vre01,
mir02}.

The continuum has a power-law slope of $\beta = -1.06 \pm 0.01$ (for
$f_\nu \propto \nu^\beta$) on the red side.  On the blue side, we find
a steeper slope of $\beta = -1.25 \pm 0.03$.  This apparent steepening
toward the blue is affected by atmospheric dispersion and is sensitive
to the centering of the OT within the slit, as well as reddening, so
the red-side value is a more reliable measurement of the intrinsic
slope.  \citet{lev02} reported a shallower optical slope of $\beta =
-0.8$ approximately 98 hours after the burst.

\section{Polarization}

Polarization was measured over five wavelength bins: $3500-4500$,
$4500-5500$, $5800-6800$, $6800-7800$, and $7800-8800$ \AA.  All
measurements were performed using the $q$ and $u$ spectra and their
associated error spectra, with the results converted to $p$ and
$\theta$ as the final step.  We observed the same three polarized
standard stars (BD+59\arcdeg389, HD 19820, and HD 155528) during both
nights of our observing run (August 12 and 13 UT).  For all three
stars and five wavelength bins, the RMS difference in the polarization
measured from the two nights' observations is only 0.04\%.  The
polarized standard stars also have essentially constant $\theta$ as a
function of wavelength (to within $\pm1\arcdeg$), and agreement to
within 1\arcdeg\ between the two nights of observations.  This
demonstrates the excellent stability of the instrument.

The raw measurements for the OT yield $p=2.3-3.1\%$ with $\theta =
153\arcdeg-162\arcdeg$, but a portion of this polarization is
caused by transmission through regions of aligned dust grains in the
Galactic ISM. The line of sight to GRB 020813 has a Galactic reddening
of $\ebv = 0.111$ mag \citep{sfd98}.  Along most Galactic lines of
sight, the maximum interstellar polarization observed is 9\% per
magnitude of $E(B-V)$ \citep{smf75}.  The ISP probe star has $p =
0.67\% \pm 0.01\%$ at $\theta = 167\fdg0 \pm 0\fdg3$ over 5800--6800
\AA, falling comfortably within the expected range, and its
$p(\lambda)$ spectrum is consistent with a Serkowski
interstellar polarization curve.  To correct the polarization
measurements of the OT, we fitted a low-order polynomial to the $q$
and $u$ spectra of the ISP star and subtracted the result from the
Stokes parameters of the OT.  The ISP-corrected measurements for the
OT are listed in Table 1; random uncertainties due to photon
statistics and detector read noise are at the level of 0.1\% in $p$
and $1-2\arcdeg$ in $\theta$.  The standard star observations suggest
that any additional systematic uncertainties should not be
time-variable, and are likely to be $\lesssim0.1\%$ in $p$ and
$\lesssim1\arcdeg$ in $\theta$.

The ISP-corrected polarization longward of 5600 \AA\ is nearly
constant during the 3 hours at 1.9--2.2\%.  On the blue side, however,
the polarization drops from $2.4\%\pm0.1\%$ during hour 1 to
$1.8\%\pm0.1\%$ in hour 3.  This is formally a $4\sigma$ detection of
polarization variability, but it is somewhat surprising to find
significant variations in $p$ only on the blue side.  As a check on
this measurement, the reductions were repeated with different
extraction and background windows, and repeated again using entirely
different reduction software; in all cases the results were consistent
with the original measurements to within 0.1\% in $p$.  We conclude
that the variability in $p$ on the blue side is likely to be genuine,
although we caution against over-interpretation of this result since
the measurement is not repeatable.

Imaging polarimetry obtained on the following night at the VLT by
\citet{cov02c} demonstrates a much more dramatic level of variability
in $p$ over the next 14 hours.  Covino \etal\ find a raw polarization
of $p=1.17\% \pm 0.16\%$ at $\theta=158\arcdeg\pm4\arcdeg$, and
ISP-corrected values of $p=0.80\%\pm0.16\%$ at
$\theta=144\arcdeg\pm6\arcdeg$.\footnote{\citet{cov02c} use an ISP
correction derived from stars in the same imaging field as the OT.
Their $V$-band ISP correction is $p = 0.59\%$ at $\theta =
178\arcdeg$, in reasonable agreement with the polarization we measure
for HD 187330.}  If we apply our ISP correction to their raw
measurement we obtain $p = 0.57\%$ at $\theta = 147\arcdeg$.  The
change in $p$ between our hour 3 data and the VLT observation is
significant at the $7\sigma$ level.  This large drop in $p$
demonstrates that most of the polarization we detect in the Keck data
must be intrinsic to the OT itself, and not merely the result of
transmission through dust within the host galaxy or the intervening
galaxy at $z=1.223$.  

The ISP-corrected polarization detected by \citet{cov02c} is
sufficiently low that in principle it could arise from the
interstellar medium in the host galaxy or perhaps even in the
intervening $z = 1.223$ system, although dust grains have a low
polarization efficiency in the ultraviolet \citep{mcw99}.
Furthermore, if GRBs result from the death of massive stars, one can
expect many GRBs to occur in dusty and gas-rich star-forming regions,
and in extreme cases Type II supernovae can have optical ISP of
several percent from the host galaxy \citep{leo02}.  However,
\emph{Chandra} observations have indicated a low absorbing column of
$N_\mathrm{H} = 7\times10^{20}$ cm\persq\ that is consistent with the
expected Galactic absorption along the line of sight \citep{vds02}.
Also, the lack of any large change in $\theta$ accompanying the
variation in $p$ suggests an intrinsic origin for the majority of the
detected polarization.  Detection of a continued late-time decline in
$p$ would provide a definitive test.

\section{Discussion}

Our observations, combined with those of \citet{cov02c}, clearly show
variability in the degree of polarization, with a nearly constant
position angle. This already places strong constraints on the
polarization models.  Random effects, such as emission from a small
number of randomly oriented magnetic field cells \citep{gw99} or
amplification of the emission from one such cell by microlensing
\citep{lp98}, would predict that $p$ and $\theta$ should change on the
same timescale.

In principle, the jet geometry, together with an assumption of some
level of orientation of the amplified magnetic field with respect to
the shock, could predict the degree of polarization as function of
time.  Models based on this scenario \citep{gl99,sari99} show that the
maximum degree of polarization occurs around the time of the jet break
(\tjet), when the opening angle of the jet is equal to the inverse
Lorentz factor.  However, at that time the emission is observable from
a significant fraction of the jet. This introduces order-unity
uncertainties in the polarization predictions since the energy
distribution profile as function of distance from the axis of the jet
is unknown. In addition, a detailed description of the hydrodynamic
evolution is not yet available; only the asymptotic scalings, well
before and well after \tjet, are understood.  Thus, it is not possible
to make detailed, quantitative comparisons of polarization models with
observations.

Despite these uncertainties, we can make general comparisons of the
observations with the model proposed by \citet{sari99}, which predicts
sharp features in the polarization light curves, although more
realistic assumptions on the hydrodynamics of jet spreading, and on
the energy distribution within the initial jet, are expected to
produce smoother features than those predicted by this model.  This
model has three parameters. The line-of-sight offset to the jet
determines the shape of the curve; \tjet\ sets the overall timescale;
and an overall scale factor for $p$ is set by a function of the ratio
of the magnetic field components parallel to and perpendicular to the
shock front, which is assumed to be constant.  We denote the offset in
units of the opening angle of the jet by $0<\phi<1$.  (For $\phi > 1$
the jet is not visible initially, corresponding to an ``orphan''
afterglow.)  Small offsets give rise to a single polarization peak,
while large offsets yield three peaks, where the middle one has
position angle perpendicular to the other two.  More realistic jet
models may smooth the curve and eliminate this middle peak.  We note
that the model of \citet{gl99} predicts only two peaks since their
calculations did not include the spreading of the jet.

Figure \ref{model} demonstrates that the jet model parameters can be
chosen so as to find at least a qualitative agreement with the data.
In the figure, our measurements for the 5800--6800 \AA\ polarization,
and the $V$-band measurement from \citet{cov02c}, are compared with
two sample models.  Initial analysis of the optical light curve
suggested $\tjet \approx 3.5-5$ hours \citep{bfh02, gh02}, although
subsequent analysis (to be presented elsewhere) indicates that the
light curves may be consistent with a later value of $\tjet \approx
10-11$ hours.  Given this uncertainty in deriving \tjet\ from the
photometric data, our approach is to determine whether there are any
values of \tjet\ that produce polarization light curves consistent
with the observations.  The first model has a small offset $\phi=0.3$
and predicts one peak of polarization. (Models with $0 < \phi
\lesssim0.35$ would give similar results.)  This requires \tjet\ to
occur during the Keck observations, at $\sim 6$ hours.  However, this
model somewhat underpredicts the decay of polarization between our
measurements and those of Covino \etal\ A better fit is obtained for a
medium-offset model with $\phi=0.55$ and $\tjet \approx 16$ hours,
where the polarization drops to zero, or perhaps even rotates the
position angle by 90\arcdeg, between the Keck and VLT observations.
Under both interpretations, the overall level of polarization is less
than the maximal possible level of $\sim20\%$.  This implies that the
magnetic field perpendicular to the shock differs only by $15-20$\%
from that parallel to the shock front.  Models with other values of
$\phi$ have difficulty matching both the near-constancy of $p$ during
the Keck observations (for $\lambda>5800$ \AA) and the large change in
$p$ detected by \citet{cov02c}.

The polarization behavior of GRB 020813 appears rather similar to that
of GRB 990712, which varied between $p=3.0\%$ at 0.4 days after the
burst and 1.2\% at $t=0.8$ days, at nearly constant position angle
\citep{rol00}.  We note that our models with $\phi=0.3$ or $0.55$
could also be nearly consistent with the GRB 990712 data within its
uncertainties.  Still, as noted by Rol \etal, the lack of clear
evidence for a jet break for that event casts some doubt on this
interpretation.

Finally, we mention that the hints of wavelength dependence to the
polarization of GRB 020813, which appear to be time-variable, can not
be accommodated by these models.  A spectral break passing through the
optical band at the time of our three measurements could produce such a
signature, but there is no evidence for such a break.

To date, polarimetric observations of GRB afterglows have had rather
sparse temporal coverage.  Further progress will require continuous
polarization monitoring of bright afterglows, initiated as early as
possible and extending well beyond \tjet.  Also, our data highlight
the need for color information in future polarization measurements.
Obtaining such observations would pose some logistical difficulties,
but the potential for improved understanding of the physical structure
of the synchrotron jets from GRBs would make this effort extremely
worthwhile.

\acknowledgments

Research by A.J.B. is supported by NASA through Hubble Fellowship
grant \#HST-HF-01134.01-A awarded by STScI, which is operated by AURA,
Inc., for NASA, under contract NAS 5-26555. R.S.  thanks the Sherman
Fairchild foundation and a NASA ATP grant for support. Data presented
herein were obtained at the W.M. Keck Observatory, which is operated
as a scientific partnership among Caltech, the University of
California, and NASA. The Observatory was made possible by the generous
financial support of the W.M. Keck Foundation.  The authors wish to
recognize and acknowledge the very significant cultural role and
reverence that the summit of Mauna Kea has always had within the
indigenous Hawaiian community.

\begin{deluxetable}{lccc}
\label{polresults}
\tablewidth{3in} 
\tablecaption{Polarization Measurements} 

\tablehead{\colhead{Hour} & \colhead{Wavelength} & \colhead{$p$} &
                          \colhead{$\theta$} \\ & \colhead{(\AA)} &
                          \colhead{(\%)} & \colhead{(\arcdeg)} }

\startdata 
1   &   3500--4500  &   $2.4 \pm  0.1$  &  $161 \pm 1$ \\
    &   4500--5500  &   $2.4 \pm  0.1$  &  $159 \pm 1$ \\
    &   5800--6800  &   $2.1 \pm  0.1$  &  $158 \pm 1$ \\
    &   6800--7800  &   $2.1 \pm  0.1$  &  $155 \pm 1$ \\
    &   7800--8800  &   $2.1 \pm  0.1$  &  $155 \pm 1$ \\
2   &   3500--4500  &   $1.8 \pm  0.1$  &  $160 \pm 1$ \\
    &   4500--5500  &   $2.0 \pm  0.1$  &  $155 \pm 1$ \\
    &   5800--6800  &   $2.0 \pm  0.1$  &  $151 \pm 1$ \\
    &   6800--7800  &   $2.0 \pm  0.1$  &  $150 \pm 1$ \\
    &   7800--8800  &   $2.1 \pm  0.1$  &  $151 \pm 2$ \\
3   &   3500--4500  &   $1.8 \pm  0.1$  &  $153 \pm 1$ \\
    &   4500--5500  &   $1.8 \pm  0.1$  &  $149 \pm 1$ \\
    &   5800--6800  &   $2.1 \pm  0.1$  &  $156 \pm 1$ \\
    &   6800--7800  &   $2.2 \pm  0.1$  &  $153 \pm 1$ \\
    &   7800--8800  &   $1.9 \pm  0.1$  &  $149 \pm 1$ \\
\enddata

\tablecomments{The results given above have been corrected for ISP
  using observations of the A2V star HD 187330, as described in the
  text.  The midpoints of the hour 1, 2, and 3 observations were at
  7:54, 9:01, and 10:06 UT, respectively.}
\end{deluxetable}

\begin{deluxetable}{lccccc}

\label{lines}
\tablewidth{4.5in} 
\tablecaption{Absorption Lines in the GRB 020813 Spectrum} 
\tablehead{\colhead{$\lambda_{\mathrm{helio}}$} & \colhead{Ion} &
  \colhead{$\lambda_{\mathrm{rest}}$} & \colhead{$z$} &
  \colhead{$W_\lambda$ (rest)} \\
  \colhead{(\AA)} & \colhead{}  & \colhead{(\AA)} &
     \colhead{} &  \colhead{(\AA)} }
\startdata 
3393.8    & \ion{Si}{2} & 1526.7 & 1.223 & $0.46 \pm 0.12$ \\
3442.3    & \ion{C}{4} blend  & 1548.2, 1550.8 & 1.222 & $2.05 \pm 0.06$ \\ 
          & +    \ion{Si}{2} & 1526.7 & 1.255 &  \\ 
3456.2    &   \nodata   & \nodata& \nodata & $1.71 \pm 0.18$\tablenotemark{a}  \\
3489.4    & \ion{C}{4}  & 1548.2 & 1.254 & $1.53 \pm 0.05$ \\
3495.1    & \ion{C}{4}  & 1550.8 & 1.254 & $1.31 \pm 0.06$ \\
3626.3    & \ion{Fe}{2} & 1608.5 & 1.254 & $1.05 \pm 0.07$ \\
3714.6    & \ion{Al}{2} & 1670.8 & 1.223 & $0.68 \pm 0.04$ \\
3766.9    & \ion{Al}{2} & 1670.8 & 1.255 & $1.48 \pm 0.04$ \\
4077.5    & \ion{Si}{2} & 1808.0 & 1.255 & $0.71 \pm 0.02$ \\
4182.6    & \ion{Al}{3} & 1854.7 & 1.255 & $0.91 \pm 0.02$ \\
4200.9    & \ion{Al}{3} & 1862.8 & 1.255 & $0.58 \pm 0.03$  \\
4569.0    & \ion{Mg}{1}, \ion{Zn}{2} , \ion{Cr}{2} & 2026.5, 2026.1, 2026.3  & 1.255 & $0.61 \pm 0.03$ \\
4636.3    & \ion{Cr}{2} & 2056.3 & 1.255 & $0.30 \pm 0.03$ \\
4650.6    & \ion{Cr}{2}, \ion{Zn}{2} & 2062.2, 2062.7 & 1.255 & $0.44 \pm 0.03$ \\
4658.5    & \ion{Cr}{2} & 2066.2 & 1.255  & $0.17 \pm 0.03$ \\
4884.2    & \nodata     & \nodata&\nodata & $0.58 \pm 0.07$\tablenotemark{a} \\
4998.5    & \ion{Fe}{2} & 2249.9  & 1.222 & $0.35 \pm 0.03$ \\
5072.5    & \ion{Fe}{2} & 2249.9  & 1.255 & $0.23 \pm 0.02$ \\
5097.3    & \ion{Fe}{2} & 2260.8  & 1.255 & $0.35 \pm 0.03$ \\
5211.9    & \ion{Fe}{2} & 2344.2  & 1.223 & $0.50 \pm 0.03$ \\
5223.2    & \nodata     &\nodata  &\nodata& $0.51 \pm 0.08$\tablenotemark{a} \\
5285.3    & \ion{Fe}{2} & 2344.2  & 1.255 & $2.00 \pm 0.02$ \\
5297.6    & \ion{Fe}{2} & 2382.8  & 1.223 & $1.01 \pm 0.03$ \\
5354.1    & \ion{Fe}{2} & 2374.5  & 1.255 & $1.42 \pm 0.03$ \\
5372.5    & \ion{Fe}{2} & 2382.8  & 1.255 & $2.10 \pm 0.03$ \\
5402.3    & \nodata     &\nodata  &\nodata& $0.64 \pm 0.07$\tablenotemark{a} \\
5752.1    & \ion{Fe}{2} & 2586.7  & 1.224 & $0.45 \pm 0.04$ \\
5782.0    & \ion{Fe}{2} & 2600.2  & 1.224 & $0.97 \pm 0.03$ \\
5811.4    & \ion{Mn}{2} & 2576.9  & 1.255 & $0.66 \pm 0.03$ \\
5833.1    & \ion{Fe}{2} & 2586.7  & 1.255 & $1.92 \pm 0.04$ \\
5851.4    & \ion{Mn}{2} & 2594.5  & 1.255 & $0.57 \pm 0.03$ \\
5863.4    & \ion{Fe}{2} & 2600.2  & 1.255 & $2.31 \pm 0.04$ \\
5877.4    & \ion{Mn}{2} & 2606.5  & 1.255 & $0.37 \pm 0.03$ \\
5882.8    & \nodata     &\nodata  &\nodata& $0.89 \pm 0.11$\tablenotemark{a} \\
5892.4    & \nodata     &\nodata  &\nodata& $0.71 \pm 0.15$\tablenotemark{a} \\
6218.4    & \ion{Mg}{2} & 2796.4  & 1.224 & $1.67 \pm 0.02$ \\
6234.3    & \ion{Mg}{2} & 2803.5  & 1.224 & $1.52 \pm 0.02$ \\
6305.7    & \ion{Mg}{2} & 2796.4  & 1.255 & $2.94 \pm 0.02$ \\
6321.9    & \ion{Mg}{2} & 2803.5  & 1.255 & $2.82 \pm 0.02$ \\
6344.4    & \ion{Mg}{1} & 2853.0  & 1.224 & $0.36 \pm 0.02$ \\ 
6433.4    & \ion{Mg}{1} & 2853.0  & 1.255 & $1.61 \pm 0.02$ \\
8872.6    & \ion{Ca}{2} & 3934.8  & 1.255 & $2.07 \pm 0.05$ \\
8951.7    & \ion{Ca}{2} & 3969.6  & 1.255 & $1.56 \pm 0.07$ \\
\enddata

\tablecomments{Observed wavelengths are heliocentric, and converted to
  vacuum.  Rest wavelengths are given as vacuum
  wavelengths. Absorption-line equivalent widths ($W_\lambda$) and
  uncertainties were measured with Gaussian fits using the IRAF SPLOT
  task, and have been transformed to the rest frame. }

\tablenotetext{a}{For unidentified absorption lines, equivalent widths
  are given in the observed frame.}
\end{deluxetable}

\begin{figure}
\begin{center}
\scalebox{0.9}{\includegraphics{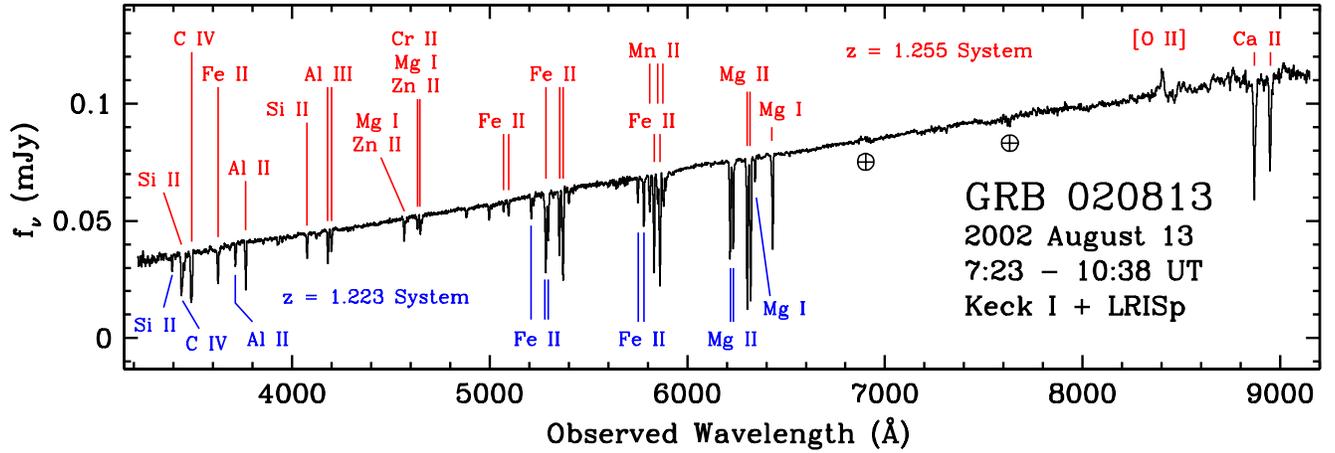}}
\end{center}
\caption{Total flux spectrum of GRB 020813, from all LRISp exposures
  combined.  Absorption features from the $z=1.255$ system are
  labelled above the spectrum, and features from the $z=1.223$ system
  are labelled below.  Residuals from the removal of the telluric A
  and B bands are marked with $\oplus$. }
\label{spectrum}
\end{figure}

\begin{figure}
\begin{center}
\scalebox{0.6}{\includegraphics{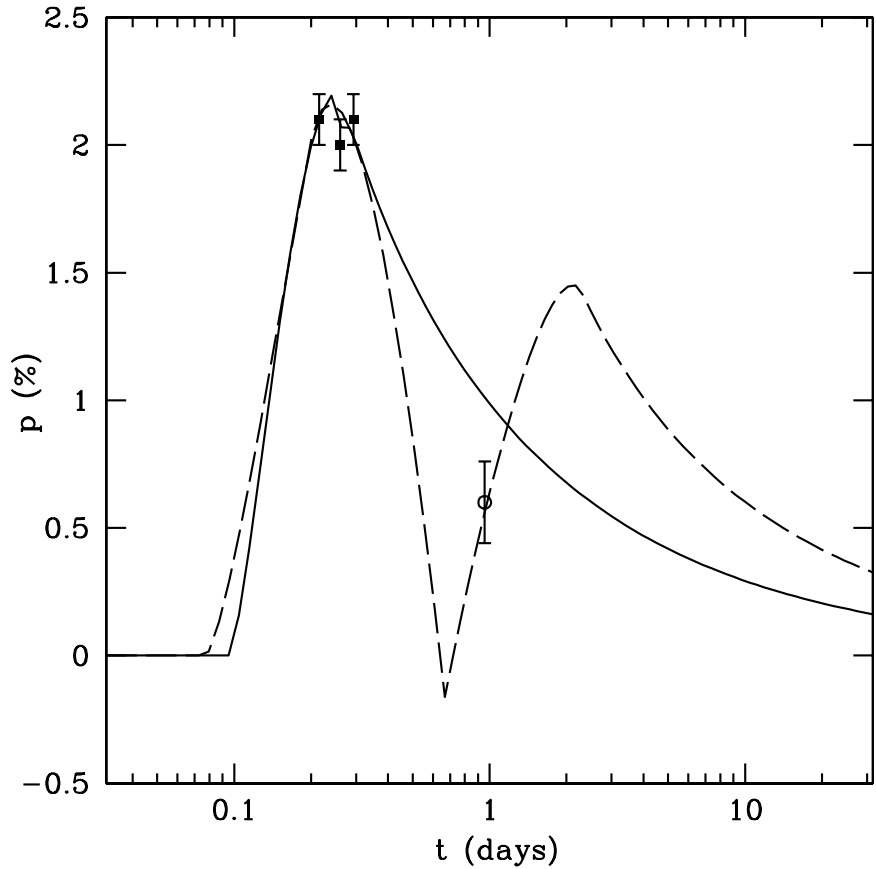}}
\end{center}
\caption{Polarization behavior of GRB 020813 compared with model
  calculations.  Squares denote the Keck observations in the
  5800--6800 \AA\ band, and the open circle is the VLT measurement
  from \citet{cov02c}.  The solid curve is the model with $\phi = 0.3$
  and $\tjet = 6$ hours, and the dashed curve is the model with $\phi
  = 0.55$ and $\tjet = 16$ hours.  Negative polarizations in this plot
  indicate a rotation of 90\arcdeg\ in the position angle relative to
  positive polarizations.  For consistency, we have applied our ISP
  correction to the VLT data point.}
\label{model}
\end{figure}

\end{document}